\documentclass[aps,physrev,twocolumn,superscriptaddress,nofootinbib,balancelastpage]{revtex4-2} 
\usepackage[latin1]{inputenc}
\usepackage{amsmath,amssymb}
\usepackage{mathrsfs} 
\usepackage{mathtools}
\usepackage{siunitx}
\usepackage{braket}
\usepackage{calc}
\usepackage{esint}
\usepackage{tabularx}
\usepackage{dsfont}
\usepackage{color}
\usepackage{ifthen}
\usepackage{graphicx}
\usepackage{seqsplit}
\usepackage[colorlinks,allcolors=black]{hyperref} 
\usepackage[capitalise]{cleveref}

\allowdisplaybreaks

\newcommand{\Eins}{\mathds{1}}%

\newcommand{\dif}{\mathrm{d}}%
\newcommand{\fdif}{\operatorname{\delta}}%
\newcommand{\Fdif}[2]{\frac{\fdif\!#1}{\fdif\!#2}}%

\newcommand{\norm}[1]{\lVert#1\rVert}%

\newcommand{\Tr}{\operatorname{Tr}}%

\newcolumntype{Y}{>{\centering\arraybackslash}X}%
\newcolumntype{Z}{>{\raggedright\arraybackslash}X}%

\newlength{\myl}%
\newcommand{\SUM}[2]{{\setlength{\myl}{\widthof{$\displaystyle\sum_{#1}^{#2}$}*\real{0.5}-\widthof{$\displaystyle\sum$}*\real{0.5}}\sum_{#1}^{#2}\;\hspace{-\the\myl}}}
\newcommand{\INT}[3]{\settowidth{\myl}{$\displaystyle\int_{#1}^{#2}$}{\int_{#1}^{#2}\;\;\;\hspace{-\the\myl}\dif #3}\,}
\newcommand{\TINT}[3]{\settowidth{\myl}{$\int_{#1}^{#2}$}{\int_{#1}^{#2}\!\ifthenelse{\equal{#1#2}{}}{}{\;\;\;\;\hspace{-\the\myl}}\dif #3}\,}%
\newcommand{\EINT}[3]{\settowidth{\myl}{$\int_{#1}^{#2}$}{\int_{#1}^{#2}\;\;\;\,\hspace{-\the\myl}\dif #3}\,}
\newcommand{\CINT}[3]{\settowidth{\myl}{$\displaystyle\int_{#1}^{#2}$}{\oint_{#1}^{#2}\;\;\;\hspace{-\the\myl}\dif #3}\,}

\begin{document}
\title{Derivation of the Thiele equation for magnetic skyrmions \\ using the Mori-Zwanzig formalism}

\author{Michael te Vrugt}
\affiliation{Institut f\"ur Physik, Johannes Gutenberg-Universit\"at Mainz, 55128 Mainz, Germany}

\author{Mathias Kl\"aui}
\affiliation{Institut f\"ur Physik, Johannes Gutenberg-Universit\"at Mainz, 55128 Mainz, Germany}

\begin{abstract}
We use the Mori-Zwanzig formalism to derive an exact generalized Langevin equation governing the dynamics of magnetic skyrmions. Using standard approximations, this generalized Langevin equation is found to reduce to the usual Thiele equation. Our approach provides a general route for obtaining extensions of the Thiele equation applicable to magnetic textures and degrees of freedom not covered by the standard derivation route. It also provides some insights into the microscopic justification of the standard Thiele equation, specifically concerning contributions of fast modes to the form of the drift and diffusion terms in the Fokker-Planck equation for skyrmions.
\end{abstract}
\maketitle

\section{Introduction}
Magnetic skyrmions \cite{ReichhardtRM2022} are particle-like spin textures with diffusive dynamics that have attracted significan attention in theoretical and experimental studies over the past years. A key reason for that is their applicability in unconventional computing \cite{MajumdarE2026,GrollierEtAl2020,EverschorSitteWM2024}. Some of these technologies \cite{RaabBBDRKMK2022,BenekeEtAl2024,BremsKV2021} explicitly rely on the thermal diffusion of the skyrmions, which consequently also has been studied more intensively \cite{ZazvorkaEtAl2019,DoiEtAl2023}. 

The dynamics of magnetic skyrmions can be described by the Thiele equation \cite{Thiele1973}, which is a Langevin equation governing the skyrmion position. It is usually derived by assuming that the time evolution of the magnetization consists solely of rigid translations. The Thiele equation has become a widely used tool in skyrmion physics \cite{EverschorGBJPF2012,MullerR2015} and allows for large-scale simulations of magnetic skyrmions on experimental length and time scales \cite{BremsSFDRKJFKV2025}, far beyond what would be computationally feasible with simulations of the microscopic theory (micromagnetism) that the Thiele equation is derived from.

Skyrmions in ferromagnets, while being among the mostly widely studied cases, are by far not the only particle-like spin texture that can be found in magnetic systems -- other examples include bimerons \cite{GobelMHMT2019,JaniEtAl2021} and the recently discovered skymeron \cite{Bhukta2026}. Relatedly, skyrmions can be found also in other magnetic phases such as antiferromagnets or altermagnets. There have, consequently, been some generalizations of the Thiele equation that include other slow modes as dynamical variables \cite{TretiakovCCBT2008} and/or that cover other phases such as altermagnets \cite{JinZCY2024,ZarzuelaJGSS2025,SouzaRSR2026} or antiferromagnets \cite{CorreiaVd2024,deSouzaCV2025}. However, these derivations usually require the assumption that the magnetization field can be expressed as a function of the slow modes. Moreover, many (though not all \cite{MiltatT2018,KamppeterMMSB1999,deSouzaCV2025}) of these works do not consider thermal noise.

A very powerful method for deriving generalized Langevin equations from a more microscopic description is the Mori-Zwanzig projection operator formalism \cite{Nakajima1958,Mori1965,Zwanzig1960,MeyerVS2017}, reviewed in Refs.\ \cite{Schilling2022,teVrugtW2019d}. It allows to derive exact equations of motion for in principle arbitrary macroscopic observables based on some known microscopic dynamics. The Mori-Zwanzig formalism is widely used in almost all fields of physics -- magnetism \cite{HartmannTGA2026,KivelsonO1974,teVrugtW2019,McHugh2016}, but also electronic structure theory \cite{KakehashiF2004}, soft matter physics \cite{AyazEtAl2021}, fluid mechanics \cite{teVrugtTWH2024}, general relativity \cite{teVrugtHW2021,Hassannejad2025}, and particle physics \cite{HuangKKR2011}. While most variants assume the microscopic dynamics to be deterministic, several authors \cite{EspanolV2002,KranzSZ2013,GlatzelS2021,Jung2026} have also considered the case of stochastic microscopic dynamics.

In this work, we present a microscopic derivation of the Thiele equation from the stochastic Landau-Lifshitz-Gilbert equation via the Mori-Zwanzig formalism. We employ for this purpose a version of the formalism developed by \citet{EspanolV2002}, which is a generalization of the formalism employed by \citet{Grabert1982}. It is applicable also to systems whose microscopic description is stochastic, allowing to treat also the noise terms governing skyrmion dynamics. Notably, our derivation route does not require the assumption that the dynamics consists solely of rigid translation, and is found to reduce to the standard result when making this assumption. Since the Mori-Zwanzig formalism allows to obtain transport equations for arbitrary observables (not just the position), it allows to generalize the Thiele formalism towards textures with orientational degrees of freedom, as required for skymerons and bimerons. Moreover, since it does not rely on the rigidity assumption employed in standard derivations of the Thiele equation, it allows to treat skyrmion dynamics with more generality.

\section{Derivation}
Our microscopic starting point is the stochastic Landau-Lifshitz equation \cite{Brown1963}
\begin{equation}
\dot{\vec{M}}= -(\vec{M}\times (\vec{H}+\vec{\xi}) - \eta (\vec{M}\times (\vec{M}\times(\vec{H}+\vec{\xi})))
\label{llg}
\end{equation}
with magnetization $\vec{M}(\vec{r},t)$ depending on position $\vec{r}$ and time $t$, effective magnetic field $\vec{H}$, damping parameter $\eta$, and noise $\vec{\xi}$ that has the properties
\begin{align}
\braket{\vec{\xi}_{\mathrm{E}}(\vec{r},t)}&=\vec{0},\\
\braket{\vec{\xi}(\vec{r},t)\otimes\vec{\xi}_{\mathrm{E}}(\vec{r}',t)} &= 2\eta k_\mathrm{B} T\Eins\delta(\vec{r}-\vec{r}')\delta(t-t')
\end{align}
with the ensemble average $\braket{\cdot}_{\mathrm{E}}$, the dyadic product $\otimes$, the unit matrix $\Eins$, the Boltzmann constant $k_\mathrm{B}$, and the temperature $T$. (We eliminate the gyromagnetic ratio by rescaling time \cite{Lakshmanan2011}.) 
The magnetization satisfies
\begin{equation}
\norm{\vec{M}}=1
\label{conservednorm}
\end{equation}
with the Euclidean norm $\norm{\cdot}$, which also implies
\begin{equation}
0=\vec{M}\cdot\partial_\nu\vec{M}.\label{funkdi2}
\end{equation}
Note that $\vec{M}$ is a three-dimensional vector, while positions $\vec{r}$ are two-dimensional (spins are located in a plane, but can rotate in three dimensions). We do not write arguments of fields unless they are unclear.

We first introduce the stochastically equivalent Fokker-Planck equation \cite{Garanin1997} 
\begin{equation}
\partial_tf[\vec{M}]= Lf[\vec{M}]
\label{fokkerp}
\end{equation}
with the Liouvillian
\begin{equation}
\begin{split}
L &= -\INT{}{}{^2r}\Fdif{}{\vec{M}}\cdot\bigg(-\vec{M}\times \vec{H} -  \eta (\vec{M}\times (\vec{M}\times\vec{H}) \\&+ \eta k_\mathrm{B}T\bigg(\vec{M}\times \bigg(\vec{M}\times \Fdif{}{\vec{M}}\bigg)\bigg)\bigg)\bigg),    
\end{split}
\end{equation}
where $f[\vec{M}]$ is to be interpreted a functional that gives the probability of finding the system in a particular configuration $\vec{M}$. As shown in Ref.\ \cite{AronBCAZL2014}, \cref{conservednorm} requires to interpret \cref{llg} in the Stratonovich calculus. Ref.\ \cite{AronBCAZL2014} also shows how to der derive \cref{fokkerp} from \cref{llg}.

The Mori-Zwanzig formalism allows, based on a microscopic description of a system (in this case given by \cref{fokkerp}), to derive equations of motion governing in principle arbitrary relevant variables, in which the other (irrelevant) variables enter as memory and noise terms. In the case that the microscopic dynamics is already stochastic, these noise terms come on top of the noise that is already present in the microscopic dynamics \cite{EspanolV2002}. The Mori-Zwanzig formalism is generally applicable, but particularly useful if the relevant and irrelevant variables correspond to the slow and fast modes of the system, respectively. Our variable of interest is the skyrmion position $\vec{R}$. It can be defined as a functional of $\vec{M}$ in the form \cite{PapanicolaouT1991,WuT2022}
\begin{equation}
\hat{\vec{R}}= \frac{1}{4\pi Q}\INT{}{}{^2r}\vec{r}\vec{M}\cdot(\partial_x \vec{M}\times\partial_y\vec{M})
\label{rasfunctional}
\end{equation}
with the topological charge $Q$. We use a hat to distinguish operators (in the sense of phase-space functions) from their measured values. A drawback of the definition \eqref{rasfunctional}, identified in Ref.\ \cite{SchutteIRN2014}, is that this expression diverges in the thermodynamic limit due to thermal fluctuations. Thus, we need to formally assume for our derivation that the system has a finite area (as it will have in an experiment/in a simulation), all spatial integrals should be interpreted as being restricted to this finite area.

With this starting point, we can perform a projection-operator based derivation following the procedure outlined in Ref.\ \cite{EspanolV2002}. We want to find the Fokker-Planck equation probability distribution $\rho(\vec{R})$ that gives the probability of finding a skyrmion at position $\vec{R}$. This probability distribution is defined as
\begin{equation}
\rho(\vec{R},t) = \INT{}{}{\vec{M}}f([\vec{M}],t)\Psi_{\vec{R}}(\vec{M}),
\label{rhodef}
\end{equation}
where the Dirac delta functional
\begin{equation}
\Psi_{\vec{R}}(\vec{M}) = \delta (\hat{\vec{R}}[\vec{M}] - \vec{R})
\end{equation}
is our relevant variable in the projection operator sense. The integral over $\vec{M}$ in \cref{rhodef} is a functional integral.

In Ref.\ \cite{EspanolV2002}, a projection operator method was developed that allows to derive a Fokker-Planck equation for a coarse-grained observable defined in terms of microscopic variables (in our case \cref{rasfunctional}) given a Fokker-Planck equation for the microscopic dynamics (in our case \cref{fokkerp}). Specializing Eqs. (2.8)--(2.14) from Ref.\ \cite{EspanolV2002} to our case, we find
the \textit{exact} result
\begin{equation}
\begin{split}
\partial_t\rho(\vec{R},t) &= \INT{}{}{^2R'}V(\vec{R},\vec{R}')\rho(\vec{R}',t) \\&+ \INT{}{}{^2R'}\INT{0}{t}{t'}K(\vec{R},\vec{R}',t-t')\rho(\vec{R}',t')
\end{split}
\label{generalfp}
\end{equation}
with the kernels
\begin{align}
V(\vec{R},\vec{R}')&=\frac{1}{\Omega(\vec{R}')}\Tr((L^\dagger\Psi_{\vec{R}})\Psi_{\vec{R}'}f_{\mathrm{eq}}),\\
K(\vec{R},\vec{R}',t)&=\frac{1}{\Omega(\vec{R}')}\Tr(f_{\mathrm{eq}}(QL^\mathrm{e}\Psi_{\vec{R}'})\exp(L^\dagger Qt)(QL^\dagger \Psi_{\vec{R}}))\label{secondkernel}
\end{align}
with the adjoint Liouvillian $L^\dagger$, the modified Liouvillian $L^\mathrm{e}$ that satisfies $LFf_{\mathrm{eq}} = f_{\mathrm{eq}}L^\mathrm{e}F$ for an arbitrary phase space function $F$, the constrained equilibrium probability
\begin{equation}
\Omega(\vec{R})=\INT{}{}{\vec{M}}\delta(\hat{\vec{R}}-\vec{R})f_{\mathrm{eq}}[\vec{M}],
\end{equation}
the equilibrium distribution $f_{\mathrm{eq}}$, the orthogonal projection operator $\mathcal{Q}=1-\mathcal{P}$, the projection operator
\begin{equation}
\mathcal{P}F=\INT{}{}{^2R}\braket{F}_{\vec{R}}\Psi_{\vec{R}},
\end{equation}
and the constrained average
\begin{equation}
\braket{F}_{\vec{R}} = \frac{1}{\Omega(\vec{R})}\INT{}{}{\vec{M}}\Psi_{\vec{R}}f_{\mathrm{eq}}[\vec{M}]F[\vec{M}].
\label{constrainedaverage}
\end{equation}

The significance of this result lies in the fact that it provides a generalized Fokker-Planck equation (or, equivalently, a generalized Langevin equation) for the skyrmion position $\vec{R}$ that is valid for any system obeying \cref{llg}. Usual derivations of the Thiele equation start from the ansatz 
\begin{equation}
M(\vec{r})=M(\vec{r}-\vec{R}),
\label{thieleansatz}
\end{equation}
and are therefore valid only insofar this ansatz holds. Thus, \cref{generalfp} constitutes a substantial generalization of the Thiele equation. Of course, its practical application requires knowledge of the kernels $V$ and $K$. These can, using established software packages and theoretical methods \cite{WidderKS2022b,MeyerWSS2021,WidderKS2022,JungHS2017}, be extracted from experimentally measured skyrmion trajectories or micromagnetic simulations. 

For evaluating $V$ we specialize Eqs. (2.16) and (2.17) from Ref.\ \cite{EspanolV2002} to our case and find
\begin{equation}
\begin{split}
V(\vec{R},\vec{R}') &=-\partial_{R_\mu}\braket{v_\mu}_{\vec{R}}\delta(\vec{R}-\vec{R}') \\&+ \frac{1}{2}k_\mathrm{B}\partial_{R_\mu}\partial_{R_\nu}\braket{d_{\mu\nu}}_{\vec{R}}\delta(\vec{R}-\vec{R}'),
\end{split}
\label{firstkernel}
\end{equation}
with
\begin{equation}
\begin{split}
v_\mu &= \INT{}{}{^2r}\vec{v}[\vec{M}]\cdot\Fdif{}{\vec{M}} \hat{R}_\mu\\&-\frac{1}{2}\eta k_\mathrm{B} T\INT{}{}{^2r}\Fdif{}{\vec{M}}\cdot \bigg(\vec{M}\times \bigg(\vec{M}\times \Fdif{}{\vec{M}}\bigg)\bigg)\bigg)\bigg)\hat{R}_\mu,
\end{split}
\label{vmu}
\end{equation}
\begin{equation}
\vec{v}[\vec{M}]=-\vec{M}\times \vec{H} -  (\vec{M}\times (\vec{M}\times\vec{H})),
\label{vecv}
\end{equation}
and
\begin{equation}
\begin{split}
d_{\mu\nu} = -\eta T\INT{}{}{^2r}(M_i M_j -\delta_{ij})\Fdif{\hat{R}_\mu}{M_i}\Fdif{\hat{R}_\nu}{M_j}.
\end{split}
\label{dmunu}
\end{equation}
We use latin indices $i=1,2,3$ for the components of $\vec{M}$ and greek indices $\mu=1,2$ for the components of $\vec{R}$. Double indices are always summed over.

For the memory kernel $K$, we make a Markovian approximation, which corresponds to assuming that all degrees of freedom except for $\vec{R}$ relax very quickly. This corresponds to (cf. Eqs. (2.18) and (2.19) in Ref.\ \cite{EspanolV2002})
\begin{equation}
\begin{split}
&\INT{0}{t}{t'}K(\vec{R},\vec{R}',t-t')\rho(\vec{R}',t')\\&\approx k_\mathrm{B}\partial_{R_\mu}\Omega(\vec{R})K_{\mu\nu}(\vec{R})\partial_{R_\nu}\frac{\rho(\vec{R},t)}{\Omega(\vec{R})}
\end{split}
\label{markova}
\end{equation}
with the kernel 
\begin{equation}
K_{\mu\nu}(\vec{R})=\frac{1}{k_\mathrm{B}}\INT{0}{\tau}{t'}\braket{(v_\nu- \braket{v_\nu}_{\vec{R}})(v_\mu - \braket{v_\mu}_{\vec{R}})},
\end{equation}
where the time scale $\tau$ is large compared to the time scales of the microscopic, but short compared to the time scales of the macroscopic degrees of freedom.

Inserting \cref{firstkernel,markova} into \cref{generalfp}, we find
\begin{equation}
\begin{split}
\partial_t\rho(\vec{R},t)&=-\partial_{R_\mu}(V_\mu(\vec{R},t)\rho(\vec{R},t))\\
&+k_\mathrm{B}\partial_{R_\mu}\partial_{R_\nu}(M_{\mu\nu}(\vec{R})\rho(\vec{R},t)).
\end{split}
\label{fphere}
\end{equation}
with the organized drift
\begin{equation}
V_\mu = \braket{v_\mu}_{\vec{R}}  + \frac{k_\mathrm{B}}{\Omega(\vec{R})}\partial_{R_\nu}(K_{\mu\nu}(\vec{R})\Omega(\vec{R})),
\label{detvel}
\end{equation}
and the mobility
\begin{equation}
M_{\mu\nu}(\vec{R}) = \frac{1}{2}\braket{d_{\mu\nu}}_{\vec{R}} + K_{\mu\nu}(\vec{R}).
\label{moblevel}
\end{equation}
This demonstrates that $V_\mu$ and $M_{\mu\nu}$ correspond to the deterministic drift and the mobility matrix appearing in the Fokker-Planck equation governing the skyrmion position. 

What remains to be shown is that the expressions that we found coincide with those in the Fokker-Planck equation corresponding to the standard Thiele equation (cf. \cite{KongZ2013,TroncosoN2014,CastellGDN2020}). For this purpose, we first use the fact that
\begin{equation}
-\vec{M}\times\vec{H}-\vec{M}\times(\vec{M}\times\vec{H}) = -\lambda\vec{M}\times\vec{H} + \lambda\eta\vec{M}\times\dot{\vec{M}}_{\mathrm{det}}
\label{llgreplacement}
\end{equation}
where $\lambda = 1/(1+\eta^2)$ \cite{Lakshmanan2011}. Here, $\dot{\vec{M}}_{\mathrm{det}}$ represents the deterministic part of the time evolution of $\vec{M}$, i.e., all terms that still remain for $T=0$. What we require is a way to write $\dot{\vec{M}}_{\mathrm{det}}$ as a functional of $\vec{M}$. Equation \eqref{thieleansatz} suggests the ansatz 
\begin{equation}
\dot{\vec{M}}_{\mathrm{det}}= -\braket{v_\alpha}_{\vec{R}}\partial_\alpha\vec{M}\label{funkdi}.
\end{equation}
This is similar to the usual assumption $\dot{\vec{M}}=-\dot{R}_\alpha \partial_\alpha \vec{M}$, but takes into account that the drift in a Fokker-Planck equation for $\vec{R}$ should not be a function of $\dot{\vec{R}}$. As discussed in Ref.\ \cite{SchmidtB2013}, quantities like $\braket{v_\alpha}_{\vec{R}}$ can play the same role as $\dot{\vec{R}}$ for the purpose of calculating averaged quantities.

From \cref{rasfunctional}, we get
\begin{equation}
\Fdif{}{\vec{M}}\hat{R}_\mu = -\frac{1}{4\pi Q}\epsilon_{\mu\nu}\vec{M}\times\partial_\nu \vec{M}
\label{fdirf}
\end{equation}
if during the variation we take into account the fact that $\norm{\vec{M}}=1$. With \cref{funkdi2}, we get
\begin{equation}
\vec{M}\cdot\Fdif{\hat{R}_\mu}{\vec{M}}=0.
\label{funkkkdi}
\end{equation}

Using \cref{vmu,vecv,llgreplacement,funkdi}, we can compute
\begin{equation}
\begin{split}
\braket{v_\mu}_{\vec{R}} &= -\frac{\epsilon_{\mu\nu}}{4\pi Q}\bigg\langle\INT{}{}{^2r}\bigg(-\lambda(\vec{M}\times\vec{H})\cdot(\vec{M}\times\partial_\nu \vec{M})\\
& +\lambda\eta(\vec{M}\times\dot{\vec{M}}_{\mathrm{det}})\cdot(\vec{M}\times\partial_\nu \vec{M})\\
&+\frac{\eta k_\mathrm{B} T}{2}\Fdif{}{\vec{M}}\cdot(\vec{M}\times(\vec{M}\times(\vec{M}\times\partial_\nu\vec{M})))\bigg)\bigg\rangle_{\vec{R}}\\
&= -\frac{\epsilon_{\mu\nu}}{4\pi Q}\bigg\langle\INT{}{}{^2r}\bigg(-\lambda((\vec{M}^2)(\partial_\nu\vec{M}\cdot\vec{H})\\& - (\vec{M}\cdot\partial_y\vec{M})(\vec{M}\cdot\vec{H}))\\
& + \lambda\eta(\vec{M}^2(\partial_\nu\vec{M}\cdot\dot{\vec{M}}) - (\partial_\nu\vec{M}\cdot\vec{M})(\vec{M}\cdot\dot{\vec{M}}))\\
& -\frac{\eta k_\mathrm{B} T}{2}\Fdif{}{\vec{M}}\cdot\vec{M}\times ((\vec{M}\cdot\partial_\nu\vec{M})\vec{M}-\vec{M}^2\partial_\nu\vec{M})\bigg)\bigg\rangle_{\vec{R}}\\
&= -\frac{\epsilon_{\mu\nu}}{4\pi Q}\INT{}{}{^2r}\bigg\langle\bigg(-\lambda(\partial_\nu\vec{M}\cdot\vec{H})\\&-\lambda\eta (\partial_\nu\vec{M})\cdot(\braket{v_\alpha}_{\vec{R}}a\partial_\alpha\vec{M})\\
&+\frac{\eta k_\mathrm{B} T}{2}\Fdif{}{\vec{M}}\cdot(\vec{M}\times\partial_\nu\vec{M})\bigg)\bigg\rangle_{\vec{R}}\\
&= \frac{\epsilon_{\mu\nu}}{4\pi Q} (F_\nu + D_{\nu\alpha}\braket{v_\alpha}_{\vec{R}}).
\end{split}
\label{vmu2}
\end{equation}
We have used the Lagrange and Gra{\ss}mann identity in second step, \cref{conservednorm,funkdi,funkdi2} in the third step, and
\begin{align}
\Fdif{}{M_k}(\epsilon_{ijk}M_i\partial_\nu M_j) &= 0,\\
F_\mu &= \lambda\braket{\INT{}{}{^2r}\partial_\mu \vec{M}\cdot\vec{H}}_{\vec{R}}
\label{fmu}
\end{align}
in the last step. Equation \eqref{fmu} should be interpreted as the definition of the force $F_\mu$. Moreover, we have defined the dissipation tensor
\begin{equation}
D_{\mu\nu}=\lambda\eta \bigg\langle\INT{}{}{^2r}(\partial_\mu\vec{M})\cdot(\partial_\nu\vec{M})\bigg\rangle_{\vec{R}}.
\label{dmdef}
\end{equation}
Since the term proportional to $T$ in \cref{vmu2} vanishes, $\braket{v_\alpha}_{\vec{R}}$ results purely from the deterministic evolution (which justifies the ansatz \eqref{funkdi}).

The effective magnetic field $\vec{H}$ can be derived from an energy functional $E$ via the equation
\begin{equation}
\vec{H}= -\Fdif{E}{\vec{M}}.
\end{equation}
The standard move is now to argue that by virtue of \cref{thieleansatz}, the energy functional $E$ becomes a function of $\vec{R}$, such that we have
\begin{equation}
F_\mu = -\partial_{R_\mu}E(\vec{R}).
\label{forcederivative}
\end{equation}
While we also employ this argument here, it is worth noting that (unlike the standard derivation route) the Mori-Zwanzig approach does not force one to do this: The force defined in \cref{fmu} is a well-defined function of $\vec{R}$ even if \cref{thieleansatz} does not hold. Moreover, writing this force in the variational form \eqref{forcederivative} can also be motivated by the fact that thermodynamic consistency (see below) requires the force to be, in the case of constant mobility, proportional to  $\partial_\mu\ln(\Omega(\vec{R}))$, which for the standard form $\Omega(\vec{R}) \propto \exp(- E(\vec{R})/(k_\mathrm{B}T))$ also leads to an expression of the form \eqref{forcederivative}.

Equation \eqref{vmu2} is a self-consistent equation for $\braket{v_\alpha}_{\vec{R}}$. We can solve it by multiplying from the left by $4\pi Q\epsilon_{\kappa\mu}$ and using
\begin{equation}
-\epsilon_{\kappa\mu}\epsilon_{\mu\nu}= \delta_{\kappa\nu},
\label{epsilon2}
\end{equation}
which leads to 
\begin{equation}
4\pi Q\epsilon_{\kappa\mu}\braket{v_\mu}_{\vec{R}}= \delta_{\kappa\nu}(F_\nu + D_{\nu\alpha}\braket{v_\alpha}_{\vec{R}}),
\end{equation}
which we can solve to get  
\begin{equation}
(4\pi \epsilon_{\kappa\mu}Q - D_{\kappa\mu}) \braket{v_\mu}_{\vec{R}} = F_\kappa.
\label{selfconsistentsolution}
\end{equation}
We now assume (as is usually done) an axisymmetric skyrmion, for which $D_{\mu\nu}=D\delta_{\mu\nu}$ \cite{CastellGDN2020}. Then, \cref{selfconsistentsolution} simplifies to 
\begin{equation}
(4\pi Q \epsilon_{\kappa\mu}- D \delta_{\kappa\mu})\braket{v_\mu}_{\vec{R}}= F_\kappa ,
\end{equation}
which has the solution
\begin{equation}
\begin{split}
\braket{v_\mu}_{\vec{R}}&= -\frac{1}{(4\pi Q)^2+D^2}(D \delta_{\mu\kappa} +4\pi Q \epsilon_{\mu\kappa})F_\kappa.
\end{split}
\label{jsolution}
\end{equation}
Next, we evaluate the diffusion tensor. With \cref{funkkkdi}, \cref{dmunu} simplifies to 
\begin{equation}
\begin{split}
d_{\mu\nu}&=\eta T\INT{}{}{^2r}\Fdif{\hat{R}_\mu}{\vec{M}}\cdot\Fdif{\hat{R}_\nu}{\vec{M}} \\&= \frac{\eta T}{(4\pi Q)^2}\INT{}{}{^2r}\epsilon_{\mu\alpha}\epsilon_{\nu\beta}(\vec{M}\times\partial_\alpha\vec{M})\cdot(\vec{M}\times\partial_\beta \vec{M})\\
&= \frac{\eta T}{(4\pi Q)^2}\INT{}{}{^2r}\epsilon_{\mu\alpha}\epsilon_{\nu\beta}(\partial_\alpha\vec{M})\cdot(\partial_\beta\vec{M}).
\end{split}
\label{dmunuintermediate}
\end{equation}
For going from the second to the third line, we have again used the Lagrange identity and \cref{conservednorm,funkdi}. Equations \eqref{dmdef} and \eqref{dmunuintermediate} imply
\begin{equation}
\braket{d_{\mu\nu}}_{\vec{R}}= \frac{T}{\lambda(4\pi Q)^2}\epsilon_{\mu\alpha}\epsilon_{\nu\beta}D_{\alpha\beta},
\end{equation}
which for an axisymmetric skyrmion simplifies to
\begin{equation}
\braket{d_{\mu\nu}}_{\vec{R}}= \frac{TD}{\lambda(4\pi Q)^2}\delta_{\mu\nu}.\label{dmunuresult}
\end{equation}

The problem with this result is that it is not compatible with the thermodynamic consistency requirement \cite{EspanolV2002}
\begin{equation}
\partial_{R_\mu} V_\mu = \partial_{R_\mu} \frac{k_\mathrm{B}}{\Omega(\vec{R})}\partial_{R_\nu} (M_{\mu\nu}\Omega(\vec{R})).
\label{thdconsistency}
\end{equation}
Such a relation between drift and dissipation term of a Fokker-Planck equation, which needs to hold at the level of the macroscopic observables if it holds at the level of the microscopic observables (which it does in our case), ensures that the system approaches a thermodynamic equilibrium distribution. Inserting \cref{detvel,moblevel} into \cref{thdconsistency} and subtracting contributions including $K_{\mu\nu}$ on both sides gives
\begin{equation}
\braket{v_\mu}_{\vec{R}}=\partial_{R_\mu} \frac{k_\mathrm{B}}{\Omega(\vec{R})}(\braket{d_{\mu\nu}}_{\vec{R}}\Omega(\vec{R})),
\end{equation}
which obviously is not compatible with \cref{jsolution,dmunuresult}.

The physical reason behind this is an inconsistency in the physical assumptions made for the drift and diffusion terms. For the drift the derivation performed here is relatively close to the standard one and uses approximations motivated by the standard ansatz \eqref{thieleansatz}. We thus make here the physical assumption that the time evolution of $\vec{M}$ consists solely of rigid translations of the skyrmion profile. While this is widely used when deriving the Thiele equation, it is not actually consistent with a microscopic model (like \cref{llg}) in which the spins experience white noise and thereby flip randomly -- randomly flipping individual spins will in general lead to a time evolution that is not just a rigid translation of a given skyrmion profile. This is why the evaluation of $\braket{d_{\mu\nu}}_{\vec{R}}$, which we were able to do, did not provide the result known from existing treatments of the stochastic Thiele equation \cite{MiltatT2018}.

What then is the physics behind these standard models? What is in fact happening in a stochastic skyrmion system is that the skyrmion is destorted by thermal fluctuations, but since the skyrmion profile is energetically favorable the system then relaxes back to this profile, but in general at a different position. If this process happens sufficiently fast then on the level of the collective variables it looks as if a rigid skyrmion is randomly moving around without deforming. This is why a derivation based on this assumption can provide a correct dynamic equation. It is plausible to assume that this is related to the known fact that the derivation of the stochastic Thiele equation does, in contrast to the derivation of the deterministic Thiele equation, not proceed without approximations even if \cref{thieleansatz} strictly holds -- at some point in the derivation one needs to replace a multiplicative by an additive noise, which provides good results but has no clear mathematical justification \cite{KamppeterMMSB1999}.

How do we fix this in the present derivation? The first observation we can make is that empirically the standard Thiele form is found to be an accurate model, which implies that the deterministic drift should in fact have the form \eqref{selfconsistentsolution}. However, as is clear from \cref{fphere,detvel}, the drift also has a contribution from the dynamics of the fast modes, namely the term proportional to $K_{\mu\nu}$. This is also reasonable based on the above physical considerations -- if the microscopic noise leads to random deviations from the rigid skyrmion profile, there has to be a deterministic force (contributing to the drift) that induces the backrelaxation. This suggests to replace \cref{selfconsistentsolution} by the more general ansatz 
\begin{equation}
(4\pi \epsilon_{\kappa\mu}Q - D_{\kappa\mu})V_\mu= F_\kappa,
\label{selfconsistentsolution2}
\end{equation}
i.e., to replace $\braket{v_\mu}_{\vec{R}}$ by $V_\mu$. This is an interesting physical result, as it implies that for a skyrmion at nonzero temperature the standard drift arises not just from rigid displacements (as the standard derivation assumes), but necessarily also has contributions from the fast modes that we have projected out. 

The total mobility is given by
\begin{align}
M_{\mu\nu}=\frac{1}{2}\frac{T D}{\lambda(4\pi Q)^2}\delta_{\mu\nu}+ K_{\mu\nu}.
\label{mmunu}
\end{align}
We can now determine $K_{\mu\nu}$. 
Using \cref{forcederivative,selfconsistentsolution2,thdconsistency,mmunu}, we find
\begin{equation}
\begin{split}
&\frac{(D \delta_{\mu\kappa} +4\pi Q \epsilon_{\mu\kappa})\partial_{R_\mu} \partial_{R_\kappa} E}{(4\pi Q)^2+D^2}\\ =&\frac{1}{2\lambda(4\pi Q)^2}D \partial_{R_\mu}^2 E + \frac{1}{T} K_{\mu\kappa}\partial_{R_\mu}\partial_{R_\kappa} E,
\end{split}
\end{equation}
which can be solved to get 
\begin{equation}
K_{\mu\nu}= T\bigg(-\frac{1}{2\lambda(4\pi Q)^2}+\frac{1}{D^2+(4\pi Q)^2}\bigg)D\delta_{\mu\nu}.
\label{kmunu}
\end{equation}
This form can be motivated by the physical interpretation of $K_{\mu\nu}$, which essentially captures the influence of stochastic deformations on the center-of-mass-motion of the skyrmion. The expression in brackets is the difference between the form of the diffusion tensor arising purely from fluctuations and the form known from the deterministic Thiele equation.

Together, \cref{mmunu,kmunu} imply 
\begin{equation}
M_{\mu\nu} = \frac{TD }{(4\pi Q)^2+D^2}\delta_{\mu\nu}.
\label{mmunufinal}
\end{equation}
Combining \cref{fphere,jsolution,mmunufinal}, we get the Fokker-Planck equation
\begin{equation}
\begin{split}
\partial_t\rho(\vec{R},t)&=-\frac{1}{(4\pi Q)^2+D^2}\partial_{R_\mu}((D \delta_{\mu\kappa} +4\pi Q \epsilon_{\mu\kappa})F_\kappa p(\vec{R},t))\\
&+ \frac{k_\mathrm{B}TD
}{(4\pi Q)^2+D^2}\partial_{R_\mu}^2 \rho(\vec{R},t).
\end{split}
\label{finalfokkerplanck}
\end{equation}
The Fokker-Planck equation \eqref{finalfokkerplanck} corresponds to the Langevin equation
\begin{equation}
\dot{R}_\nu=\frac{1}{(4\pi Q)^2+D^2}(D \delta_{\nu\kappa} +4\pi Q \epsilon_{\nu\kappa})(F_\kappa + \chi_\kappa),
\label{almostthiele}
\end{equation}
where the noise $\chi_\mu$ has the properties
\begin{align}
\braket{\chi_\mu(t)}_{\mathrm{E}}&=0,\\
\braket{\chi_\mu(t)\chi_\nu(t')}_{\mathrm{E}}&= 2k_\mathrm{B}TD  \delta_{\mu\nu}\delta(t-t').
\end{align}
If we multiply \cref{almostthiele} from the left with $D \delta_{\mu\nu} + 4\pi Q \epsilon_{\nu\mu}$, we get
\begin{equation}
4\pi Q \epsilon_{\nu\mu}\dot{R}_\nu + D \dot{R}_\mu =  F_\mu +\chi_\mu,
\end{equation}
which is the Thiele equation.

\section{Discussion}
The advantages of this approach are:
\begin{itemize}
\item It does not rely on the validity of the ansatz \eqref{thieleansatz}, but works generally and allows to compute correction terms.
\item The formalism in principle also gives a way to extract these correction terms from experimental data.
\item Already for the derivation of the standard Thiele equation, it provides new insights into the justification of standard approximations, in particular regarding the origin of the noise term (whose treatment in the standard derivation, in particular concerning the multiplicative noise, raises some questions \cite{KamppeterMMSB1999}).
\item Moreover, one can extend it to other variables (such as internal modes of a skyrmion, orientation of a bimeron etc.) in order to obtain Thiele equations for these variables as well.
\end{itemize}
The disadvantage is that it is in most cases not obvious how to evaluate the kernels analytically.

Moreover, our results are of theoretical interest for researchers studying the Mori-Zwanzig formalism:
\begin{itemize}
    \item The derivation of the Thiele equation from the Landau-Lifshitz-Gilbert equation presented here is in a way reminiscent of the derivation of the Friedmann equations in cosmology from the Einstein field equations in Ref.\ \cite{teVrugtHW2021}, where the Mori-Zwanzig formalism was extended to general relativity. Also in the present work, one uses this formalism to derive an ordinary differential equation from a partial differential equation. Most derivations \cite{EspanolL2009,WittkowskiLB2012,EspanolV2002,AneroET2013} proceed the other way round.
    \item Moreover, the present work shows a derivation in which one uses the Mori-Zwanzig formalism to derive a Langevin equation from another Langevin equation, whereas most applications take Hamiltonian dynamics as a microscopic starting point.
    \item Finally, it is somewhat interesting that we were not able to derive the correct form of the Thiele equation using solely the organized drift. The reason this is notable is that in the derivation of the Dean equation from Brownian dynamics presented in Ref.\ \cite{EspanolV2002} this \textit{was} possible, which is ultimately a consequence of the fact that the Dean equation is an exact representation of the Brownian dynamics it is derived from. Usual derivations of the Thiele equation would suggest that this is also possible since once the ansatz \eqref{thieleansatz} is made we do not require further approximations to obtain it. However, in the stochastic case we need to make approximations to evaluate the noise term. 
\end{itemize}

\section{Conclusion}
In this work, we have presented a derivation of the Thiele equation from the stochastic Landau-Lifshitz-Gilbert equation using the Mori-Zwanzig formalism, which establishes that the Thiele equation can be interpreted as a generalized Langevin equation in the projection operator sense. This allows, among other things, to see how the diffusion behavior of thermal skyrmions is affected by fast modes. An obvious extension, for which this work is intended as a starting point, is to develop extensions of the Thiele model for other magnetic textures -- such as skymerons and bimerons -- for which additional (orientational) degrees of freedom need to be incorporated. On the theoretical side, it would be of interest to provide a fully microscopic justification of \cref{selfconsistentsolution2,kmunu} that does not require an appeal to the fluctuation-dissipation theorem. Finally, taking into account the non-Markovian elements in the generalized Langevin equation (which we mostly neglected in the present work) will allow to systematically incorporate, e.g., the coupling to magnonic heat baths \cite{WeissenhoferRN2021}. Previous studies \cite{PsaroudakiHKL2017,PsaroudakiAL2019} indicate that this coupling in general leads to non-Markovian friction terms, which is in line with what the generalized Langevin equation approach predicts.

\acknowledgments{We thank Hartmut L\"owen, Achim Rosch, Abhinav Sharma, and Peter Virnau for helpful discussions. M.t.V.\ is funded by the Deutsche Forschungsgemeinschaft (DFG, German Research Foundation) -- SFB 1552, Project-ID 465145163.}

\end{document}